\documentclass[12pt]{iopart}
\usepackage[dvips]{graphicx,epsfig}
\begin{document}

\title[]{Neutral Pions with Large Transverse Momentum 
in {\it d}+Au and Au+Au Collisions}

\author{Henner B\"usching
\footnote[3]{buschin@bnl.gov}
 for the PHENIX Collaboration}

\address{Brookhaven National Laboratory, Upton, NY 11973-5000, USA}

\begin{abstract}
 Measurements of transverse-momentum ($p_T$) spectra 
 of neutral pions in Au+Au and {\it d}+Au collisions at 
 $\sqrt{s_{\mathrm{NN}}}=200~\mathrm{GeV}$ and $62.4~\mathrm{GeV}$
 by the PHENIX experiment at RHIC in comparison
 to {\it p+p} reference spectra at the same $\sqrt{s_{\mathrm{NN}}}$ are
 presented.
 In central Au+Au collisions at $\sqrt{s_{\mathrm{NN}}}=200~\mathrm{GeV}$
 a factor $4-5$
  suppression for neutral pions and charged hadrons with $p_T >
  5~\mathrm{GeV}/c$ is found relative to the {\it p+p} reference
  scaled by the nuclear overlap function $\langle T_\mathrm{AB} \rangle$. In
  contrast, such a suppression of high-$p_T$ particles is absent in
  {\it d}+Au collisions independent of the centrality of the collision.
  To study the $\sqrt{s_{\mathrm{NN}}}$ dependence of the suppression
  Au+Au collisions at 
 $\sqrt{s_{\mathrm{NN}}}=200~\mathrm{GeV}$ and $62.4~\mathrm{GeV}$
  are compared.

\end{abstract}




\section{Introduction}

The heavy-ion program at the Relativistic Heavy Ion Collider (RHIC) 
at the Brookhaven National Laboratory is exploring 
the fundamental theory of strong interactions, QCD, under extreme conditions. 
A main goal is to create and study a deconfined and thermalized 
state of strongly interacting matter, the quark-gluon plasma (QGP). 
Most particles created in central Au+Au collision at RHIC 
carry rather low transverse momenta ($p_T < 2~\mathrm{GeV}/c$). 
For the understanding of nucleus-nucleus (A+A) collisions, however,
the small fraction of particles with high $p_T$
originating from parton-parton interactions with high momentum transfer
are of particular importance.
In A+A collisions the hard
parton-parton scatterings happen in the early stage of the collision,
before a possible QGP has formed. The high-energetic
scattered partons traverse the subsequently produced excited nuclear
matter. As a consequence, particle production at high $p_T$
is sensitive to properties of the hot and dense matter
created in A+A collisions. 

Here we study high-$p_T$ neutral pions at mid-rapidity which are measured with
the two central spectrometer arms of the PHENIX experiment. Each arm
covers $|\eta| \le 0.35$ in pseudorapidity and $\Delta \phi = \pi/2$
in azimuth. Neutral pions were measured by the PHENIX electromagnetic
calorimeters (EMCal) via the $\pi^0 \rightarrow \gamma \gamma$ decay.
The EMCal consists of six lead-scintillator and
two lead-glass sectors, each located at a radial distance of
about 5~m to the interaction region~\cite{Aphecetche:zr}.
The systematic uncertainty of the absolute energy scale of the
EMCal is 1.5\% in this measurement.

\section{High-$p_T$ Particle Yields in Au+Au collisions at $\sqrt{s_{\mathrm{NN}}}=200~\mathrm{GeV}$}

Hadron production mechanisms in A+A are usually studied
via their scaling behavior with respect to {\it p+p} collisions. 
Soft processes ($p_{T}<$ 1 GeV/$c$) are expected to
scale with the number of participating nucleons $N_{part}$.
Hard parton-parton interactions with small cross section,
however, can be considered as an incoherent sequence of individual 
nucleon-nucleon collisions.
In the absence of any medium effects the production of high-$p_T$ 
particles should be comparable to the
production in {\it p+p} after scaling with a geometrical factor which reflects
the increased number of scattering centers.
It is customary to quantify the medium effects at high $p_{T}$ 
using the {\it nuclear modification factor} which is
given by the ratio of the A+A to the {\it p+p} invariant yields~\cite{Adler:2003pb}
that are scaled by the 
{\sl nuclear overlap function} $\langle T_{\mathrm{AB}}\rangle$:
\begin{equation}
R_{AA}(p_{T})\,=\,\frac{d^2N^{\pi^0}_{AA}/dy dp_{T}}{\langle T_{\mathrm{AB}}\rangle\,\times\, d^2\sigma^{\pi^0}_{pp}/dy dp_{T}}.
\label{eq:R_AA}
\end{equation}
The average nuclear overlap function $\langle T_\mathrm{AB}\rangle$ is
determined solely from the geometry of the nuclei A and B. The average 
number of nucleon-nucleon collisions per
A+B collision is given by $\langle N_{\mathrm{coll}}\rangle =
\sigma_{\mathrm{inel}}^\mathrm{pp} \times \langle
T_\mathrm{AB}\rangle$.

\begin{figure}
\begin{center}
\epsfig{file=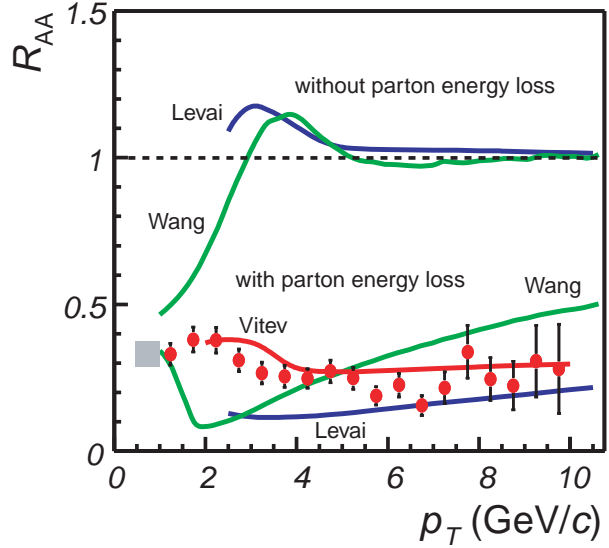,width=8cm}
\end{center}
\caption{Nuclear modification factor $R_\mathrm{AA}(p_T)$ 
for neutral pions in central Au+Au collisions at $\sqrt{s_{\mathrm{NN}}}=200~\mathrm{GeV}$
compared to theoretical predictions as discussed in the text.}
\label{fig_AuAu}
\end{figure}

$R_{\mathrm{AA}}(p_{T})$ measures the deviation of A+A from an incoherent superposition
of $NN$ collisions in terms of suppression ($R_{\mathrm{AA}}<$1) or enhancement ($R_{\mathrm{AA}}>$1).
In contrast to the naive expectation the results for central Au+Au collisions 
at $\sqrt{s_{\mathrm{NN}}}=200~\mathrm{GeV}$ at RHIC
show a suppression of up to a
factor of five in the nuclear modification factor
\cite{Adler:2003qi,Adler:2003au} as shown in Fig. \ref{fig_AuAu}.

\begin{figure}
\begin{center}
\epsfig{file=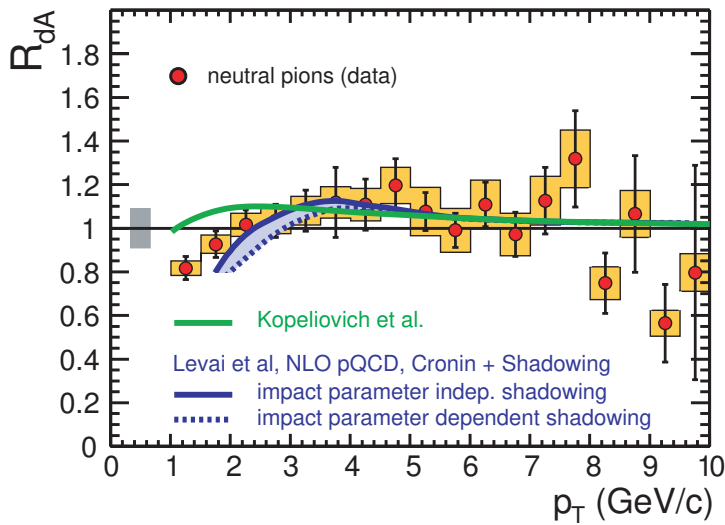,width=10cm}
\end{center}
\caption{Nuclear modification factor $R_\mathrm{dA}(p_T)$ 
for neutral pions in {\it d}+Au collisions at $\sqrt{s_{\mathrm{NN}}}=200~\mathrm{GeV}$
compared to theoretical predictions. The boxes around the data points indicate the 
systematic errors correlated in $p_T$, 
and the shaded grey band at 1 indicates the percent normalization error.}
\label{fig_dAu}
\end{figure}

It has been suggested that the observed suppression is a result of
parton energy loss in a medium of high color-charge density (mainly
gluons) like the quark-gluon plasma
\cite{Gyulassy:1990ye,Wang:2003yp}. 
A number of model predictions (shown in Fig.~\ref{fig_AuAu})
which were made prior to the release of the data can describe the 
data if they take into account constant (Wang et.al.)~\cite{pre_constantE} or
energy-dependent (Levai et. al.)~\cite{pre_nonconstantE1}, (Vitev et. al.)
~\cite{pre_nonconstantE2} parton energy-loss effects.
Although the magnitude of the suppression in the data at $p_T \sim 4$~GeV/c
is qualitatively described by the model with constant energy loss, 
the increasing $R_{\mathrm{AA}}$ with $p_T$ is contrary to what
is observed in the data; whereas the energy-dependent energy-loss models 
give reasonable agreement with the data.

In contrast to parton energy loss descriptions, a gluon saturation calculation~\cite{dima}
is able to predict the magnitude of the observed suppression as well, 
though it fails to reproduce exactly the flat $p_{T}$ dependence of the quenching.
This model assumes a saturation of gluons at small Bjorken-$x$ in the initial-state
wavefunction of the Au nuclei, which leads to fewer hard gluon-gluon
scatterings and thus to fewer high-$p_T$ particles.
Similarly, estimates of final-state interactions in a dense {\it hadronic}
medium~\cite{gallmeister} yield the same amount of quenching as models based on partonic
energy loss.
So the latter two models claim to explain the observed suppression without assuming 
the formation of a quark-gluon plasma.

Hence it is impossible to distinguish between initial-state effects (like gluon saturation)
and final-sate effects (like parton energy loss) by studying solely Au+Au collisions at this energy.

\section{High-$p_T$ Particle Yields in {\it d}+Au collisions at $\sqrt{s_{\mathrm{NN}}}=200~\mathrm{GeV}$}

\begin{figure}
\begin{center}
 \epsfig{file=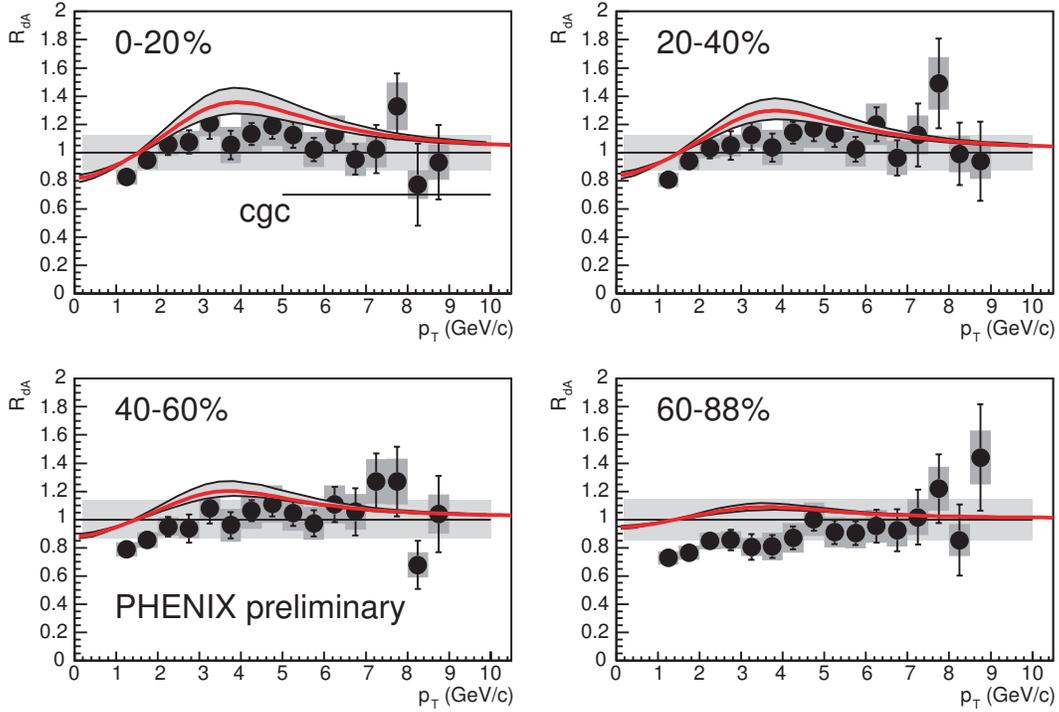,width=14cm}
\end{center}
\caption{Nuclear modification factor $R_\mathrm{dA}(p_T)$ 
for neutral pions in {\it d}+Au collisions at $\sqrt{s_{\mathrm{NN}}}=200~\mathrm{GeV}$
for four different centrality selections compared to theoretical calculations.
}
\label{fig_dAu_cent}
\end{figure}

Explanations attributing the high-$p_T$ hadron suppression in central Au+Au collision to 
final-state effects and
models which are based on initial-state effects make different predictions for {\it d}+Au collisions:
In a {\it d}+Au collision the highly excited matter is not created over
a large volume and so high-$p_T$ particle suppression is
not expected in the parton energy-loss scenario.  By contrast, in
the framework of the gluon saturation model high-$p_T$ particle
production ought to be suppressed in {\it d}+Au as well. The gluon
saturation calculation in \cite{dima} predicts $R_\mathrm{AB} \approx
0.7$ for central {\it d}+Au collisions. {\it d}+Au reactions have been studied in early
2003 by PHENIX and first results have been published \cite{Adler:2003ii}.

\begin{figure}
\begin{center}
 \epsfig{file=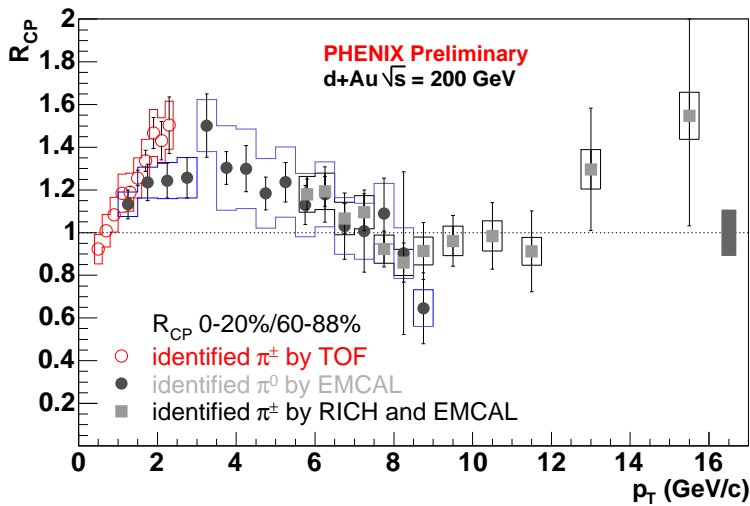,width=10cm}
\end{center}
\caption{Comparison of pion yields
in the most central $(0-20\%)$ and the most peripheral $(60-88\%)$ data samples $R_\mathrm{CP}$
for three different measurements as a function of $p_{T}$.}
\label{fig_pion}
\end{figure}

The nuclear modification factor $R_{dA}$ for high-$p_T$ $\pi^{0}$'s 
in minimum-bias {\it d}+Au collisions is shown in Fig.~\ref{fig_dAu}.
In contrast to central Au+Au collisions no suppression at mid-rapidity can be seen.
Within uncertainties $R_{dA}$ for neutral pions is equal to unity in contrast
to unidentified charged particles that exhibit a clear enhancement compared to {\it p+p} collisions
(not shown). This enhanced production, observed first in $p+A$ fixed-target 
experiments (``Cronin effect'')~\cite{cronin}, is interpreted in terms 
of multiple initial-state soft and semi-hard interactions which 
broaden the transverse momentum of the colliding partons prior to 
the hard scattering.
Figure~\ref{fig_dAu} also shows that models incorporating both the Cronin enhancement
and shadowing effects can describe the data well:
A next-to-leading-order pQCD prediction which was made prior to the release of the data 
(Levai et. al.~\cite{levai}) with impact-parameter-dependent and independent shadowing 
effects and a model with a different implementation of Cronin and shadowing effects
(Kopeliovich et. al.~\cite{Kopeliovich:2002yh}) are compared to the data.
These {\it d}+Au results support the conclusion that a final-state nuclear medium effect, 
such as parton energy loss, is necessary to describe the suppression observed in central
Au+Au collisions.

The influence and interplay of various nuclear effects on particle production
depends on the geometry or centrality of the collision
due to the changing nuclear density.
Particle production at different rapidities also reflects the dynamics of the 
nuclear and Bjorken-$x$ dependence of these effects.  
Therefore, a careful study of centrality- and pseudorapidity-dependence of 
particle production in {\it d}+Au
collisions can provide an important baseline for the understanding of
the relative contributions of different initial and final-state effects
in Au+Au collisions. Here we will focus on the centrality dependence
of neutral pion production - studies of pseudorapidity dependence can be found in~\cite{Liu:2004kg}.
Figure \ref{fig_dAu_cent} shows the nuclear modification factor $R_\mathrm{dA}(p_T)$ 
for neutral pions in {\it d}+Au collisions at $\sqrt{s_{\mathrm{NN}}}=200~\mathrm{GeV}$
for four different centrality selections compared to theoretical calculations.
The boxes around the data points indicate the systematic errors correlated in $p_T$, 
and the shaded grey band at 1 indicates the percent normalization error.
Within systematic errors $R_\mathrm{dA}(p_T) \approx 1$. There is a hint for a small
increase of $R_\mathrm{dA}$ with centrality. This would be in accordance with the
observation for unidentified charged spectra which show a clear 
increase of $R_\mathrm{dA}$ 
with centrality in {\it d}+Au, a trend opposite to the Au+Au results.

The data can be well described over all centralities by a Glauber-Eikonal approach~\cite{Accardi:2003jh}
shown as the red curve in Fig.~\ref{fig_dAu_cent} without any assumption about initial-state
gluon-saturation effects. The prediction for the gluon-saturation calculation (CGC, \cite{dima})
is plotted as well. Even with the systematic uncertainties of the preliminary
PHENIX data the model prediction seems to underestimate the data in the most central data
sample.

To demonstrate the consistency of various measurements of charged and neutral pions 
within the PHENIX experiment Fig.~\ref{fig_pion} shows the ratio $R_\mathrm{CP}$
of pion yields
in the most central $(0-20\%)$ to the most peripheral $(60-88\%)$ data samples
each normalized by the respective $\langle T_\mathrm{AB} \rangle$
for three different measurements as a function of $p_T$. 
This ratio has the advantage that many of the systematic 
errors cancel. The figure shows measurements of  charged pions at
low $p_T$ identified by the TOF detector~\cite{Adcox:zp} and at high $p_T$ 
identified by the RICH and EMCal detectors together with the measurement of
$\pi^{0}$'s. All three independent measurements agree well within systematic errors.
In future analyses it has to be proven if $R_\mathrm{CP}$ for 
$p_T > 8~\mathrm{GeV}$ stays at one.

\section{High-$p_T$ Particle Yields in Au+Au collisions at $\sqrt{s_{\mathrm{NN}}}=62.4~\mathrm{GeV}$}

\begin{figure}
\begin{center}
 \epsfig{file=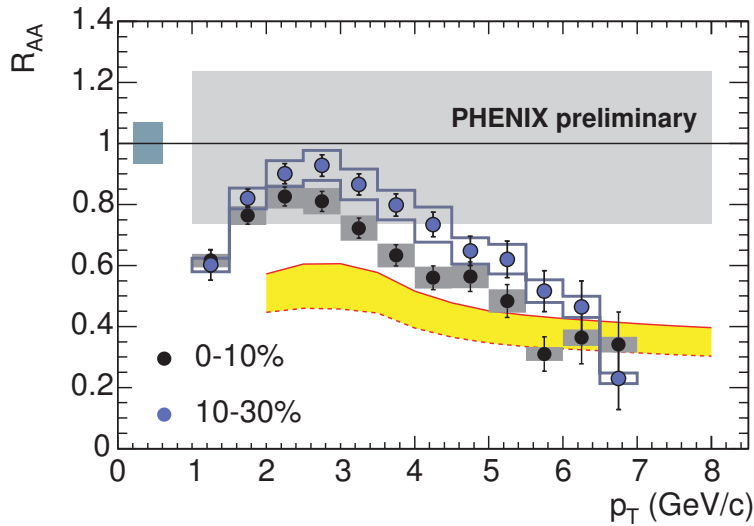,width=10cm}
\end{center}
\caption{Nuclear modification factor $R_\mathrm{AA}(p_T)$ 
for neutral pions in Au+Au collisions at $\sqrt{s_{\mathrm{NN}}}=62.4~\mathrm{GeV}$
for two different centrality selections compared to a theoretical prediction.}
\label{fig_62}
\end{figure}

To study the onset of the suppression of neutral pions 
observed in central Au+Au collisions 
at $\sqrt{s_{\mathrm{NN}}}=200~\mathrm{GeV}$ and to map out the elastic and inelastic 
scattering properties of a possibly created quark-gluon plasma it is useful to 
study the $\sqrt{s_{\mathrm{NN}}}$ dependence of  neutral-pion production.
To provide more information in the energy range
between the measurements at CERN SPS~\cite{wa98_pi0} at 
$\sqrt{s_{\mathrm{NN}}}=17.3~\mathrm{GeV}$ and the RHIC measurements at 
$\sqrt{s_{\mathrm{NN}}}=130~\mathrm{GeV}$ and 200 GeV in early 2004
the RHIC experiments measured particle production in Au+Au collisions
at $\sqrt{s_{\mathrm{NN}}}=62.4~\mathrm{GeV}$.
Figure \ref{fig_62} shows the results on 
the nuclear modification factor $R_\mathrm{AA}(p_T)$ 
for neutral pions in Au+Au collisions at this energy
for two different centrality selections.
The boxes around the data points indicate the systematic errors correlated in $p_T$, 
and the shaded grey band at 1 indicates the percent normalization error.
Note the huge systematic error on the normalization due to large uncertainties in
the {\it p+p} reference as there is no $\pi^0$ measurement from RHIC
available at this energy~\cite{denterria}.
$R_\mathrm{AA}(p_T)$ for $\pi^0$'s shows a suppression of up to a factor 5 at high $p_T$
with a strong $p_T$ dependence.
A model prediction which was made prior to the release of the data~\cite{Vitev:2004gn}
incorporating the final-state energy loss with a gluon
rapidity density of $dN^{g}/dy=650-800$
and initial-state Cronin scattering
is shown in Fig.~\ref{fig_62} as well.
For an interpretation of the data a future measurement of the {\it p+p} reference
at RHIC at this energy is strongly desirable.

\section{Conclusions}

One of the most significant observations at RHIC so far is the suppression of
high-$p_T$ hadrons at mid-rapidity in central $\sqrt{s_{\mathrm{NN}}}=200~\mathrm{GeV}$ 
Au+Au collisions relative to 
binary-scaled {\it p+p} reference spectra. This observation
is consistent with predictions in which hard-scattered partons lose energy
in a dense medium. A competing theory, the gluon saturation model, appears
as an unlikely explanation after the same measurements were made in {\it d}+Au
collisions: Here no high-$p_T$ particle suppression is observed at mid-rapidity
supporting the expectation of the parton energy loss scenario but disproving 
predictions by the gluon saturation model. Very little centrality dependence 
of $R_\mathrm{AA}(p_T)$ for neutral pions is observed in {\it d}+Au.
The $\sqrt{s_{\mathrm{NN}}}$ dependence is studied by $\pi^0$ spectra in Au+Au collisions
at $\sqrt{s_{\mathrm{NN}}}=62.4~\mathrm{GeV}$.
\\

\end{document}